\documentclass[aps,superscriptaddress, showpacs,preprintnumbers, superscriptaddress, nofootinbibt,twocolumn]{revtex4}
\usepackage{eurosym}
\usepackage{amsfonts}
\usepackage{amssymb,amsmath,enumitem}
\usepackage{graphicx,color,epstopdf}
\usepackage{subfigure}

\def\be{\begin{equation}}
\def\ee{\end{equation}}
\def\bea{\begin{eqnarray}}
\def\eea{\end{eqnarray}}

\begin{document}

\title{Comment on ``Reexamining $f\left(R,T\right)$ gravity'',  Phys. Rev. D 100, 064059 (2019)}
\author{Tiberiu Harko}
\email{tiberiu.harko@aira.astro.ro}
\affiliation{Astronomical Observatory, 19 Ciresilor Street,  Cluj-Napoca 400487, Romania,}
\affiliation{Department of Physics, Babes-Bolyai University, Kogalniceanu Street,
Cluj-Napoca 400084, Romania,}
\affiliation{School of Physics, Sun Yat-Sen University, Xingang Road, Guangzhou 510275,
P. R. China,}
\author{Pedro H. R. S. Moraes}
\email{moraes.phrs@gmail.com}
\affiliation{Departamento de F\'{i}sica, Instituto Tecnol\'{o}gico de Aeron\'{a}utica, Centro
T\'{e}cnico Aeroespacial, 12228-900 S${\rm \tilde{a}}$o Jos\'{e} dos Campos, S${\rm \tilde{a}}$o Paulo, Brazil}

\begin{abstract}
In a recent paper, ``Reexamining $f\left(R,T\right)$ gravity'', by S. B.
Fisher and E. D. Carlson, Phys. Rev. D 100, 064059 (2019), the authors claim
that for the particular $f(R,T)$ modified gravity model, with $%
f(R,T)=f_1(R)+f_2(T)$, the term $f_2(T)$ must be included in the matter
Lagrangian and therefore it does not have any physical significance. We
carefully reexamine the line of reasoning presented in the paper, and we show that
there are several major conceptual problems related to the author's physical interpretations, as well as in the physical and mathematical approaches used to derive the energy-momentum tensor of the theory. These problems raise some serious concerns about the validity of most of the results presented in the paper.
\end{abstract}

\pacs{03.75.Kk, 11.27.+d, 98.80.Cq, 04.20.-q, 04.25.D-, 95.35.+d}
\maketitle

\tableofcontents

\section{Introduction}

In an interesting and thought provoking paper, ``Reexamining $%
f\left(R,T\right)$ gravity'' by S. B. Fisher and E. D. Carlson \cite{1}, the
authors propose a mathematical, as well as a physical reformulation of a
specific version of the $f(R,T)$ gravity theory \cite{2}, in which the
gravitational Lagrangian $f(R,T)$ can be decomposed as $f(R,T)=f_1(R)+f_2(T)$. Here $R$ is the Ricci scalar, $T$ denotes the trace of the energy-momentum tensor and $f_1(R)$ and $f_2(T)$ are functions specifically dependent on $R$ and $T$, respectively. In the present Comment, we carefully reexamine the results of the paper \cite{1} and we show that there are several conceptual problems in their physical analysis and interpretation of the $T$-dependence of the $f(R,T)$ gravity.

We will first briefly summarize in the following what the authors
of \cite{1} are essentially doing.

Let us begin with the gravitational action
\begin{eqnarray}
S&=&\int{\left[\frac{1}{2\kappa ^2}f(R,T)+L_m\right]\sqrt{-g}d^4x}  \notag
\\
&=&\int{\left[\frac{1}{2\kappa ^2}f_1(R)+f_2(T)+L_m\right]\sqrt{-g}d^4x},
\end{eqnarray}
where $\kappa^ 2=8\pi G$, $f_1(R)$ and $f_2(T)$ are arbitrary (analytical) functions of $R$ and $T$, respectively,  $L_m$ is the matter action, and $g$ is the metric determinant. Since both $f_2(T)$ and $L_m$ are
functions of the same argument, that is, of the thermodynamical parameters of the
system, one can obviously combine them into a single term $L_m^{(\text{eff})}$,
\begin{equation}  \label{1}
L_m^{(\text{eff})}=f_2(T)+L_m,
\end{equation}
so that the gravitational action of the $f(R,T)$ theory takes an \textit{%
effective} form, given by
\begin{equation}  \label{1a}
S=\int{\left[\frac{1}{2\kappa ^2}f_1(R)+L_m^{(\text{eff})}\right]\sqrt{-g}d^4x}.
\end{equation}

Now the main \textit{physical claim} of the authors of \cite{1} is that the
effective matter Lagrangian (\ref{1}) is the true matter Lagrangian of the
system and that ``$f_2(T)$ is not physically meaningful''. But, before going
into the discussion of the claimed physical usefulness of $f_2(T)$, let us consider
firstly the simple example of the scalar field, as discussed in \cite{1}.

\section{The case of the scalar field}

For a scalar field $\phi$ with potential $V(\phi)$, the Lagrangian $L_{\phi}$
and its trace $T_{\phi}$ are (in the notations of \cite{1})
\begin{equation}  \label{3}
L_{\phi}=\frac{1}{2}\left[\nabla _{\mu}\phi \nabla ^{\mu} \phi-V(\phi)\right]%
, T_{\phi}=-\nabla _{\nu}\phi \nabla ^{\nu}\phi+2V(\phi).
\end{equation}

Then, in the framework of the $f(R,T)=f_{1}(R)+f_{2}(T_{\phi})$ theory, a scalar
field Lagrangian of the form
\begin{equation}
L_{\phi }^{({\text{eff}})}(\phi )=L_{\phi }+\alpha \ln \left\vert T_{\phi
}\right\vert ,  \label{4}
\end{equation}%
where $\alpha $ is a constant, is perfectly justified.

Let us analyze now the
possibility of reducing this scalar field Lagrangian to its standard form (%
\ref{3}). For this we consider a general transformation of the scalar field
given by $\phi =F(\Phi )$, where $F$ is an arbitrary function of a new scalar field $\Phi$. Since we have
\be
\nabla _{\mu }\phi =\frac{ dF(\Phi
)}{d\Phi } \nabla _{\mu }\Phi ,
\ee
the scalar field Lagrangian becomes
\begin{eqnarray}  \label{5}
L_{\Phi }(\Phi )&=&\frac{1}{2}\left[\frac{dF\left( \Phi \right) }{d\Phi }%
\right] ^{2}\left[ \nabla _{\mu }\Phi \nabla ^{\mu }\Phi -U\left( \Phi
\right) \right]  \notag \\
&+&\alpha \ln \left\vert -\nabla _{\nu }\Phi \nabla ^{\nu }\Phi +2U(\Phi
)\right\vert +2\alpha \ln \frac{dF\left( \Phi \right) }{d\Phi },  \notag \\
\end{eqnarray}
where $U\left( \Phi \right) =V(\Phi )/\left[ dF(\Phi )/d\Phi \right] ^{2}$.
Hence Eq. (\ref{4}) is almost form invariant with respect to an arbitrary
transformation of the scalar field and therefore the claim
``For more complicated functions $f_{2}(T)$, the resulting terms will of
course not be simply a rescaling of the field, but will change the free
field into an interacting field.'' in \cite{1} does not look to be true ({\it in a natural way})
for arbitrary functions $f_2\left(T_{\phi}\right)$.

\section{The energy momentum tensor in $f(R,T)=f_1(R)+f_2(T)$ gravity}

Next we need to clarify the concept of matter, ``physical pressure'' and ``physical energy density'', respectively. By matter we usually understand a
system of interacting particles, whose structure and dynamics are determined
by the known (or lesser known) laws of nature. Based on these laws, the
physical parameters of fluids, like the four-velocity, for example, and the
set of thermodynamic scalars, such as energy density, pressure, temperature and
specific enthalpy, can be defined uniquely in an instantaneous Lorentz
frame carried by the fluid, and determined accordingly experimentally.

 However, in order to have a correct understanding of the terms we are using in our Comment it is necessary to explain the definitions of physical and effective thermodynamic quantities. In a broad sense, we mean by physical quantities those defined in standard textbooks of physics, like, for example, \cite{Lan}. In a more restricted sense {\it we define the physical (thermodynamical)  quantities  as the quantities that are obtained from the microscopic distribution functions of the particles (Fermi-Dirac, Bose-Einstein, Boltzmann etc.)}.

 The presence of the gravitational field modifies the distribution function at the microscopic level, and at the level of the total energy. The problem of the gravitational energy and of its localization is a complex one (not yet considered in $f(R,T)$ gravity), but once we succeed in including it in the distribution functions of the particles, we can obtain from them the thermodynamic potentials, energy density, pressure etc.. The quantities obtained in this way indeed correspond to the ``true'' physical and thermodynamical variables in a gravitational field. However, they will depend essentially on the metric, and any role played by $f_2(T)$, if any,  is uncertain. For a discussion of the problem of the energy-momentum pseudotensor of the gravitational field in standard general relativity see \cite{Lan1}.

 On the other hand we can construct thermodynamic like quantities by simply combining (additively) the physical pressure with other similar quantities of different origins. We call these kind of quantities {\it effective quantities}, and they are not ``true'' physical quantities in the sense previously defined, since (generally) {\it they cannot be derived from microscopic distribution functions of particles}.

Hence the scalar physical thermodynamic quantities {\it cannot be rescaled arbitrarily by adding to
them some functions of the same thermodynamic parameters}. One cannot claim
that the pressure of the degenerate fermionic relativistic gas, given by $%
p\propto \rho ^{4/3}$ is not physical, and cannot be determined
experimentally, and that the true pressure of the degenerate Fermi gas is
(let us say), $p\propto \rho ^{4/3}+f_2(T=\rho-3p)$. {\it One should clearly point
out that the standard thermodynamic quantities (density, temperature,
pressure etc.) are real physical quantities that can be obtained from the microscopic distribution function of particles, and as such, they are at the
theoretical foundations for the description of gravitational processes
involving the presence of matter}. On the other hand the thermodynamic quantities considered in \cite{1} can be considered only {\it effective quantities}, and they are definitely not the ``true'' physical pressures or energy densities of any (real) physical system, since they cannot be obtained from any (known) classical or quantum distribution function of particles.

\subsection{Constructing the energy-momentum tensor}

In order to extract some useful information from the action (\ref{1a}), and
to construct an energy-momentum tensor for the theory, the authors of \cite%
{1} adopt the formalism developed in \cite{3}, by using for the matter
Lagrangian density the standard expression (16) in their paper. Then the
authors arrive at the ``on shell'' current densities and particles number,
defined in Eqs. (21a) and (21b), which include the function $f_2(T)$ and
which are conserved ``on shell''. However, since in this approach {\it the physical
current $J^{\mu}$ and the physical particle number $n$ are not conserved},
the authors reach the controversial conclusion that in this model there may be
some ``true'' energy density, given by their Eqs. (27) and (29), and which also may
represent the experimentally measurable energy density.

The problem of the construction of the energy-momentum tensor in modified
theories of gravity with geometry-matter coupling was discussed in \cite{4}
and \cite{5} (see also \cite{6}).

However, in the present Comment we use a different, and more physical approach \cite{7,8}. First we require that the variations of
the entropy density $s$ and of the ordinary matter number flux vector
density,
\be
n^{\mu }=nu^{\mu }\sqrt{-g},
\ee
where $n$ is the particle number, defined as
\be
n=\sqrt{\frac{g_{\mu \nu }n^{\mu }u^{\nu }}{g}},
\ee
 satisfy the constraints
 \be
 \delta s=0,
 \ee
 and
 \be
 \delta n^{\mu }=0,
 \ee
 respectively, which maintain unchanged the entropy and particle production rates. Therefore, the entropy and particle number currents satisfy the conservation equations $\delta \left(
n^{\mu }\partial _{\mu }s\right) =0$ and $\nabla _{\mu }\left( nu^{\mu
}\right) =0$, respectively.

Let the equation of state for matter be given as $\rho =\rho \left(
n,s\right) $. Then, since $\delta s=0$, from the thermodynamic relation $%
\left( \partial \rho /\partial n\right) _{s}=w=\left( \rho +p\right) /n$, we
obtain $\delta \rho =w\delta n$. By taking the variation of the particle
number $n$ we find \cite{8}
\begin{eqnarray}
\delta n&=&\frac{n}{2}\left( -g\right) u^{\mu }u^{\nu }\left( \frac{\delta
g_{\mu \nu }}{g}-\frac{g_{\mu \nu }}{g^{2}}\delta g\right) =  \notag \\
&&-\frac{n}{2}\left( u^{\mu }u^{\nu }+g^{\mu \nu }\right) \delta g_{\mu \nu
}.
\end{eqnarray}

For the sake of concreteness, and for simplicity,  we assume that the ordinary matter Lagrangian
is $L_{m}=-\rho $ \cite{8}, and we introduce the effective matter action as
\begin{equation}
S_{m}^{(eff)}=-\int \left[ \rho -f_{2}\left( T\left( n,s\right) \right) %
\right] \sqrt{-g}d^{4}x.
\end{equation}

By taking the variation of $S_{m}^{(eff)}$, we obtain
\begin{eqnarray}
\delta S_{m}^{(eff)} &=&-\int \Bigg\{\left[ \delta \rho -\frac{df_{2}(T)}{dT}%
\frac{dT\left( n,s\right) }{dn}\delta n\right] \sqrt{-g}+  \notag \\
&&\left[ \rho -f_{2}\left( T\left( n,s\right) \right) \right] \delta \sqrt{-g%
}\Bigg\}d^{4}x=  \notag \\
&&-\int \left[ w-\frac{df_{2}(T)}{dT}\frac{dT\left( n,s\right) }{dn}%
\right] \delta n\sqrt{-g}-  \notag \\
&&\frac{1}{2}\left[ \rho -f_{2}\left( T\left( n,s\right) \right) \right]
\sqrt{-g}g^{\mu \nu }\delta g_{\mu \nu },
\end{eqnarray}%
immediately giving
\begin{eqnarray}
&&\hspace{-0.5cm}\delta S_{m}^{(eff)}=\frac{1}{2}\int \Bigg\{\left[ \rho +p-n\frac{df_{2}(T)%
}{dT}\frac{dT(n,s)}{dn}\right] u^{\mu }u^{\nu }+  \notag \\
&&\hspace{-0.5cm}\left[ p+f_{2}\left( T\left( n,s\right) \right) -n\frac{df_{2}(T)}{dT}%
\frac{dT(n,s)}{dn}\right] g^{\mu \nu }\Bigg\}\delta g_{\mu \nu }\sqrt{-g}%
d^{4}x.  \notag \\
\end{eqnarray}

Hence the {\it effective matter energy-momentum tensor} that can be constructed
from the action (\ref{1a}) is given by
\bea
^{(eff)}T^{\mu \nu }&=&\left[ \rho +p-n\frac{df_{2}(T)}{dT}\frac{dT(n,s)}{dn}%
\right] u^{\mu }u^{\nu }+\nonumber\\
&&\left[ p+f_{2}\left( T\left( n,s\right) \right) -n%
\frac{df_{2}(T)}{dT}\frac{dT(n,s)}{dn}\right] g^{\mu \nu }, \nonumber\\
\eea
and it obviously reduces to the perfect fluid form of standard general
relativity when $f_{2}(T)=0$. $^{(eff)}T^{\mu \nu }$ can also be written in
terms of {\it an effective energy density $\rho _{eff}$ and pressure $p_{eff}$},
defined as
\begin{equation}
\rho _{eff}(n,s)=\rho (n,s)-f_{2}\left( T\left( n,s\right) \right) ,
\end{equation}
\begin{equation}
p_{eff}(n,s)=p(n,s)+f_{2}\left( T\left( n,s\right) \right) -n\frac{df_{2}(T)%
}{dT}\frac{dT(n,s)}{dn},
\end{equation}%
in the form
\be
^{(eff)}T^{\mu \nu }=\left( \rho _{eff}+p_{eff}\right) u^{\mu }u^{\nu
}+p_{eff}g^{\mu \nu }.
\ee

This energy-momentum tensor is different from the one obtained in \cite{1} on both conceptual and mathematical levels. For the fluid description of matter, and in standard general relativity, {\it the same expression for the energy-momentum tensor can be obtained if one assumes for the matter Lagrangian the expression $L_m=p$, and by decomposing the velocity in terms of scalar potentials} \cite{8}. If the degeneracy of the matter Lagrangian can be removed in $f(R,T)$ gravity, due to the presence of the matter-geometry coupling, and if yes, how this can be done, is a problem that goes beyond the topics of the present Comment, and thus we will not consider it.

We must also point out that in the present approach {\it the ordinary matter satisfies all the standard conservation laws}, without the necessity of introducing any ``on shell'' quantities, and conservation laws, or modifying the physical interpretation of the ordinary thermodynamical quantities.

\section{Conclusions}

We can now summarize our main findings as follows. The claim in \cite{1} that the function $f_2(T)$ can be just simply included in the matter action in  a {\it physical} way is questionable. The matter energy density and  pressure are two fundamental thermodynamic quantities that are obtained from microscopic particle distribution functions, and they cannot be arbitrarily modified without completely changing the content of the physical laws, and of the corresponding theories. However, such a construction is perfectly valid from the point of view of the construction of {\it effective} physical quantities.  The mathematical/physical approach employed by the authors of \cite{1} to derive the energy-momentum tensor of the $f(R,T)=f_1(R)+f_2(T)$ gravity theory, {\it even correct mathematically}, raises some questions about its physical interpretation, since matter satisfies the conservation equations of the current and entropy only ``on shell'' (that is, they are not conserved in the true physical sense). The ``true'' physical quantities describing matter in any physical theory are $\rho $ and $p$, and they are not equivalent in any sense (be it mathematical or physical) with {\it the effective quantities that also include $f_2(T)$}. Hence the search for the functional form of $f_2(T)$ is a valid one, and finding observational restrictions/constraints on the function $f_2(T)$, as done, for example, in \cite{9} and \cite{10}, is an important field of research that could lead to some new insights and a better understanding of the mathematical structure and astrophysical and cosmological implication of the $f(R,T)$ gravity.

\section*{Acknowledgments}

T. H. would like to thank the Yat-Sen School of the Sun Yat-Sen University in Guangzhou, P. R. China, for the kind hospitality offered during the preparation of this work. P. H. R. S. M. would like to thank São Paulo Research Foundation (FAPESP), grant 2015/08476-0, for financial support. He also thanks the financial support of FAPESP under the thematic project 2013/26258-4.

\end{document}